\documentclass[letterpaper]{aa}  
\usepackage{graphicx}
\usepackage{longtable,lscape}
\usepackage{lscape}
\usepackage{txfonts}
\begin{document}
\title{Chemical behavior of the  Dwarf Irregular Galaxy NGC\,6822.\\ Its PN and \ion{H}{ii}  region abundances\thanks{Based on observations collected at the European Southern Observatory, VLT, Paranal, Chile, program ID 077.B-0430} \thanks{Based on observations obtained at the Gemini Observatory (program ID G-2005B-56), which is operated by AURA, Inc. under a cooperative agreement with the NSF on behalf of the Gemini partnership: the National Science Foundation (USA), the Science and Technology Facilities Council (United Kingdom), the
National Research Council (Canada), CONICYT (Chile), the Australian Research Council
(Australia), Minist\'erio da Ciencia e Tecnologia (Brazil) and SECYT (Argentina).}}
\author{Liliana Hern\'andez-Mart\'inez\inst{1}, Miriam Pe\~na\inst{1}, Leticia Carigi\inst{1} and Jorge Garc\'ia-Rojas\inst{1}%\thanks{On sabbatical leave  at the Departamento de Astronom{\'\i}a, Universidad de Chile.}
}
\offprints{L. Hern\'andez-Mart\'inez}
\institute{Instituto de Astronom{\'\i}a, Universidad Nacional Aut\'onoma de M\'exico, Apdo. P ostal 70264, M\'ex. D. F., 04510 M\'exico\\
\email{lhernand,miriam,carigi,jgarcia@astroscu.unam.mx}
 }
\date{Received 15/05/2009; Accepted  13/06/2009}
\titlerunning{Chemical Abundances in NGC6822}
\authorrunning{Hern\'andez-Mart\'inez et al.}


\abstract
{}
{ We aim to derive the chemical behavior of a  significant sample of PNe and HII regions in the irregular galaxy NGC\,6822  The selected objects are distributed in different zones of the galaxy. Our purpose is
to obtain the chemical abundances of the present interstellar medium (ISM), represented by \ion{H}{ii} regions and the corresponding values
at the time of formation of  PNe. With these data the chemical homogeneity of NGC\,6822
will be tested and the abundance pattern given by \ion{H}{ii} regions and PNe will be used as an
observational constraint for computing chemical evolution models to infer the chemical history
of NGC 6822.}
{Due to the faintness of PNe and \ion{H}{ii} regions in NGC\,6822, to gather spectroscopic data  with large telescopes  is  necessary.  We obtained a well suited sample of spectra by employing VLT-FORS\,2 and Gemini-GMOS spectrographs. Ionic and total abundances are calculated for the objects  where electron temperatures can be determined through the detection of [\ion{O}{iii}]\,$\lambda$4363 or/and [\ion{N}{ii}]\,$\lambda$5755 lines. A ``simple'' chemical evolution model has been developed and the observed data are used to compute a model for NGC\,6822 in order to infer a preliminary chemical history in this galaxy. }
{Confident determinations of He, O, N, Ne, S and Ar abundances were derived for a sample of 11 PNe and one \ion{H}{ii} region. We confirm that the present ISM is chemically homogeneous, at least in the central 2 kpc of the galaxy, showing a value 12\,+\,log O/H = 8.06$\pm$0.04.  From the abundance pattern  of PNe, we identified two populations: a group of  young PNe  with abundances similar to \ion{H}{ii} regions and a group of older objects with abundances a factor of two lower.  A couple of extreme Type I PNe were found. No third dredge-up O enrichement was detected in PNe of this galaxy. The abundance determinations  allow us   to discuss the chemical behavior of the present and past ISM in NGC\,6822. Our preliminary  chemical evolution model predicts that an important gas-mass lost  occurred during the first 5.3 Gyr, that no star higher than 40 M$_\odot$ was formed, and
that   1\% of all 3-15 M$_\odot$ stars became binary systems progenitors to SNIa.} 
{}
\maketitle
\section{Introduction}

Dwarf Irregular Galaxies are the ``simplest'' galactic systems known. They are considered one of the building blocks of the Universe and apparently they are dominated by dark matter (Carigi et al. 1999). Two of their main characteristics are their low mass and low metallicity. They are gas-rich galaxies and thought to be chemically homogeneous. Even when they are simple in appearance, they are dominated by star forming complexes and OB  associations with typical diameters of 200-300 pc (Fisher \& Tully 1979; Hodge et al. 1991) which are immersed in an older and more extended populations.

NGC\,6822 (DDO 209, IC 4895) is one of the closest gas-rich dwarfs in the Local Group. The galaxy appears very extended on the sky due to its small distance from the Milky Way, it's optical apparent dimensions are about 15.5$'$ $\times$13.5$'$, and it is at a distance modulus of 23.31$\pm$0.02 as reported by  Gieren et al. (2006). It is moving at $V_{\rm hel}= -54 \pm 6$ km s$^{-1}$ and presents an optical radius  of 2.9 kpc $\times$ 2.9 kpc (Mateo 1998). A huge and rotating  H\,{\sc i}  disk of about 6 kpc  $\times$ 14 kpc size at P.A. $\sim$ 110$^\circ$ is a well known feature in this galaxy (de Blok \& Walter 2000 and references therein). Its young stellar content extends in zones with radii over 5 kpc from the center (de Blok \& Walter 2006) and it also shows an even larger distribution of faint C stars forming a spheroid whose long axis lies almost perpendicular to the H\,{\sc i} disk (e.g., Demers et al. 2006). NGC\,6822 was believed to be isolated, but recent studies argue the presence of a ``North West" companion (de Blok  \& Walter 2000,  2006). This galaxy has a total luminosity of M$_{\rm B}=-15.8$ (Hodge et al. 1991) and a total H\,{\sc i} mass of 1.3\,$\times$\,10$^8 $M$_\odot$ (de Blok \& Walter 2006). 
 
On the chemical context NGC\,6822 is a metal-poor galaxy, with an interstellar medium (ISM) abundance of about 0.2 Z$\odot$ (Richer \& McCall 2007). Its seems to be chemically homogeneous, as expected for dwarf irregulars. Lee et al. (2006) tested the existence of a possible radial gradient with no conclusive results. So far, a few bright \ion{H}{ii}  regions have been studied by different authors. From collisionally excited lines, Hidalgo-G\'amez et al. (2001) reported 12\,+\,log(O/H)\,=\,8.10 and 8.12 for the regions Hubble V (H\,V) and Hubble X (H\,X); Peimbert et al. (2005) re-analyzed the same regions deriving similar values from collisionally excited lines, while from recombination lines they obtained 12\,+\,log(O/H)\,=\,8.42. The latter value is more in agreement with the values reported for A type supergiants (Venn et al. 2000). 

Regarding planetary nebulae (PNe), due to their faintness only a couple of spectra were analyzed in the past (Dufour \& Talent 1980; Richer \& McCall 1995). Recently, Richer \& McCall (2007) obtained spectroscopic data of a sample of seven PNe. They reported that PNe in NGC\,6822 have oxygen and neon abundances very similar to those in the \ion{H}{ii} regions.

It has been generally accepted that  $\alpha$-elements like oxygen and neon are not affected significantly by the nucleosynthesis processes occurred during the evolution of low-intermediate mass stars (LIMS), progenitors of PNe, which on the other hand do modify He, N and C. Therefore,  O and Ne of PNe would reflect the metallicity of the ISM from which the parent stars were formed and their abundances were usually taken as representative  of the ISM in that epoch. Recently,  some observational evidences and theoretical models have shown that this is questionable; 
  O and Ne original abundances in PNe can be perturbed  by stellar nucleosynthesis in at least two ways. In more massive progenitors O can be depleted through ON-cycle (e.g., Peimbert 1985; Henry 1990; Costa et al. 2000). On the other hand, low metallicity PN progenitors can dredge up freshly synthesized  O (and possibly also Ne, besides C) to their surfaces, in the third dredge-up episode  (Marigo 2001;  Herwig 2004; Leisy \& Dennefeld 2006; Wang \& Liu 2008), although this apparently happens significantly at very low metallicity. For instance,  in some very poor galaxies it has been found that oxygen abundances of PNe (and apparently also Ne) are higher than the average value of H II regions. Such is the case in the only PN known in Sextans A (Kniazev et al. 2005; Magrini et al. 2005), in the 8 PNe studied in NGC\,3109 by Pe\~na et al. (2007), in a large fraction  of PNe in the SMC and in a small but significant fraction in the  LMC (Leisy \& Dennefeld 2006).
Thus, neither O nor Ne  can be considered safe for testing the original chemical composition of progenitor stars.

Our main goal in this work is to analyze the chemical  abundances of a larger sample of PNe in NGC\,6822 and to compare them with the present ISM abundances. In this galaxy there are 26 PN candidates detected (Hern\'andez-Mart{\'\i}nez \& Pe\~na 2009, hereafter Paper I) from which we selected a sample for spectroscopic analysis. Based on well measured electron temperatures T(\ion{O}{iii}) we determine the chemical abundances  of 11 PNe and one \ion{H}{ii}  region and analyze their  abundance pattern.  These data are used to compute one ``simple'' chemical evolution model for NGC\,6822 in order to infer a preliminary chemical history in this galaxy. In a future paper (Hern\'andez-Mart{\'\i}nez et al. 2009, in preparation), we will present a more detailed grid of models and a deeper discussion on this subject.

      The paper is organized as follows: In \S 2 we present the observations and data reduction. Analysis of the spectra and plasma diagnostic are presented in \S3. The behavior of chemical abundances is discussed in \S4 and in \S5 the history of chemical enrichment in NGC\,6822 is analyzed by means of a model of chemical evolution. Our conclusions are presented in \S6.
      
   \section{Observations and data reduction}
  To accurately determine nebular abundances, Êdata in a large wavelength range, Êat Êresolution large enough to easily separate important diagnostic lines such [\ion{O}{iii}] $\lambda$4363 from H$\gamma$ and the components of the doublet [\ion{S}{ii}]  $\lambda\lambda$6717,6731,  are required. Also exposure times  should be long enough to obtain the faint diagnostic lines (usually about hundreds of times fainter that H$\beta$) with good signal-to-noise. In particular to measure the temperature sensitive  [\ion{O}{iii}] 4363/5007 ratio  is crucial.
 
 With this purpose, we obtained two set of observations with large aperture telescopes by employing multi-object spectrographs (MOS) in order to observe simultaneously a large number of PNe and compact \ion{H}{ii}  regions. First, a service-mode run with the Gemini South telescope and the GMOS spectrograph (2 fields centered at 19:44:56 -14:42:58 and 19:44:59  -14:48:08, each with size  5.5$\times$5.5 arcmin, were observed) was performed on 2005 which allowed us to gather high-quality data for a  couple of PNe and a few \ion{H}{ii}  regions (program ID GS-2005B-Q-56). A second run with the ESO Very Large Telescope UT1 (Antu) and the FORS2 spectrograph, in MXU mode (covering one field centered at 19:44:56.7 -14:47:57.0,  6.8$\times$6.8 arcmin size) and also in long-slit mode, was performed on 2006 August  20 and 21. Both, GMOS and FORS2-MXU,  employ masks constructed from previously acquired images (pre-imaging). The log of observations is presented in Table 1. The objects for spectroscopy were selected from the lists of PN candidates and \ion{H}{ii}  regions reported by Leisy et al. (2005) and in   Paper I. The objects are designed as in Paper I, where the reader can find  the cross-correlation with other names in the literature.
 
 Gemini South GMOS pre-imaging were acquired on August 2005, and used to select objects for spectroscopy. Spectroscopic observations were performed  through  gratings B600 and R600 with a spectral resolution of 0.45 and 0.47 \AA/pix respectively. The covered  wavelength range depends on the position of the object in the field of view but in general the range from  4000 \AA ~ to 7500 \AA~ was obtained. Slit width was of 1$''$ for all the objects, but of varying lengths to accommodate a larger number of objects. In this case, the exposure times were not long enough for the fainter objects whose faint lines  were  underexposed. Therefore, from these observations we are reporting only a couple of  PNe and three faint \ion{H}{ii}  regions which have good enough signal-to-noise for plasma analysis. The standard stars LTT\,9239 and H\,600 were used for flux calibration.
 
  For the spectroscopic VLT-FORS2 run, the grisms 600B and 600RI were used. Again, the spectral  range coverage depends  on the position of the object, but   in general it goes from about 3700  $\AA$ to 7500 $\AA$ with a spectral resolution of 0.7 to 1.2 \AA/pix. The slit width was 1$''$ for all the objects. The standard stars EG\,274, LDS\,749B and BMP\,16274 were observed through a slit of 5$"$ width for flux calibration.  During the spectroscopic run the sky was clear and the seeing conditions varied from 0.7$"$ to 0.9$"$.
  
 In both set of observations, the spectral resolution is good enough to safely separate  lines such as [\ion{O}{iii}] 
 $\lambda$4363   from H$\gamma$ and the doublet [\ion{S}{ii}] $\lambda$$\lambda$6717, 6731, which are important for plasma diagnostics.

\begin{table*}
\begin{center}
\footnotesize{
\caption{Log of observations }
\begin{tabular}{lrrr}
\hline \hline 
Objects$^1$& spectrum mode & \multicolumn{2}{c}{grism (exp.time (s)$\times$ \# exposures)} \\
\hline
\multicolumn{1}{c}{VLT FORS2}\\
\hline
PN\,4, PN\,6 &Long slit & 600B  (1500$\times$3)& 600RI  (1500$\times$2) \\
PN\,10, PN\,12, PN\,14, PN\,16, PN\,18 & MXU           & 600B   (1800$\times$3) &600RI  (1500$\times$3) \\
PN\,19, PN\,21, \ion{H}{ii} 15, \ion{H}{ii} 9         \\
\hline
\multicolumn{1}{c}{Gemini South}\\
\hline
N: PN\,5,  PN\,7, \ion{H}{ii}  2, \ion{H}{ii}  4, \ion{H}{ii}  5  &GMOS& B600 (1100$\times$4) & R600$^2$ (900$\times$4)\\
%S: &GMOS& B600 & 550$\times$4\\
\hline \hline
\multicolumn{4}{l}{$^1$Names from the list by Hern\'andez-Mart{\'\i}nez \& Pe\~na (2009).} \\
\multicolumn{4}{l}{$^2$ R600 was used with the filter $GG455\_G0329$.}
\end{tabular}
\label{tab:obsvlt}
}
\end{center}
\end{table*}

 \subsection{ Data reduction,  spectral calibrations, and reddening correction}

Normal data reduction was performed by using IRAF\footnote{IRAF is distributed by the National Optical Astronomy Observatories, which is operated by the Association of Universities for Research in Astronomy, Inc., under contract to the National Science Foundation.}  reduction packages. Raw frames were bias subtracted and flat-fielded.  Wavelength calibration was performed with a He-Hg-Cd lamp for FORS2 data and a Cu-Ar lamp for GMOS data. Flux calibration was performed via the standard stars mentioned above.

Line fluxes were measured with the IRAF routine `onedspec.splot', by applying a Gaussian profile fit. For the case of VLT-FORS2 observations, the red and blue  spectra of each object were scaled to a common level using lines detected in both spectral ranges.The logarithmic reddening correction at H$\beta$ was obtained from the Balmer decrement. Spectral lines were dereddened with the Seaton extinction law (Seaton 1979). The dereddened fluxes are presented in Table 2 where the reddening law is also listed. For the objects observed with Gemini-GMOS (PN\,5, PN\,7 and HII\,4) such a procedure cannot be performed because the blue and red spectra do not overlap, they were observed in different days under different conditions, and the flux calibration is not quite confident. Thus the Balmer line ratios, H$\alpha$/H$\beta$ and H$\gamma$/H$\beta$ indicate very different reddening corrections. As it is very important for abundance determination, to use the intensity of [\ion{O}{iii}] $\lambda$4363 and other lines as reliable as possible, we used observed H$\alpha$/H$\beta$ and H$\gamma$/H$\beta$ ratios and we calculated the additional corrections to bring them to the theoretical recombination values. Such a correction was then applied  to the observed flux at other wavelengths.

It is interesting to notice that the objects located in the center, near or on  the ``galactic bar'', such as \ion{H}{ii}\,15, \ion{H}{ii}\,9, PN\,10  present higher reddening --in average E(V-B)$\sim$0.54-- than the objects located in the periphery, such as (PN\,14,PN\,16,PN\,18,PN\,19) which in average show E(B-V)$\sim$0.25. This is in agreement with previous results indicating a
reddening varying from E(B-V)=0.24 in the border of the galaxy to E(B-V)=0.45 in the center (Massey et al. 1995).

Our spectra of PNe and one \ion{H}{ii} region are shown in Fig. 1.
\begin{figure*}[ht] 
\begin{center}
\label{spectra}
\includegraphics[width=17cm,height=17cm]{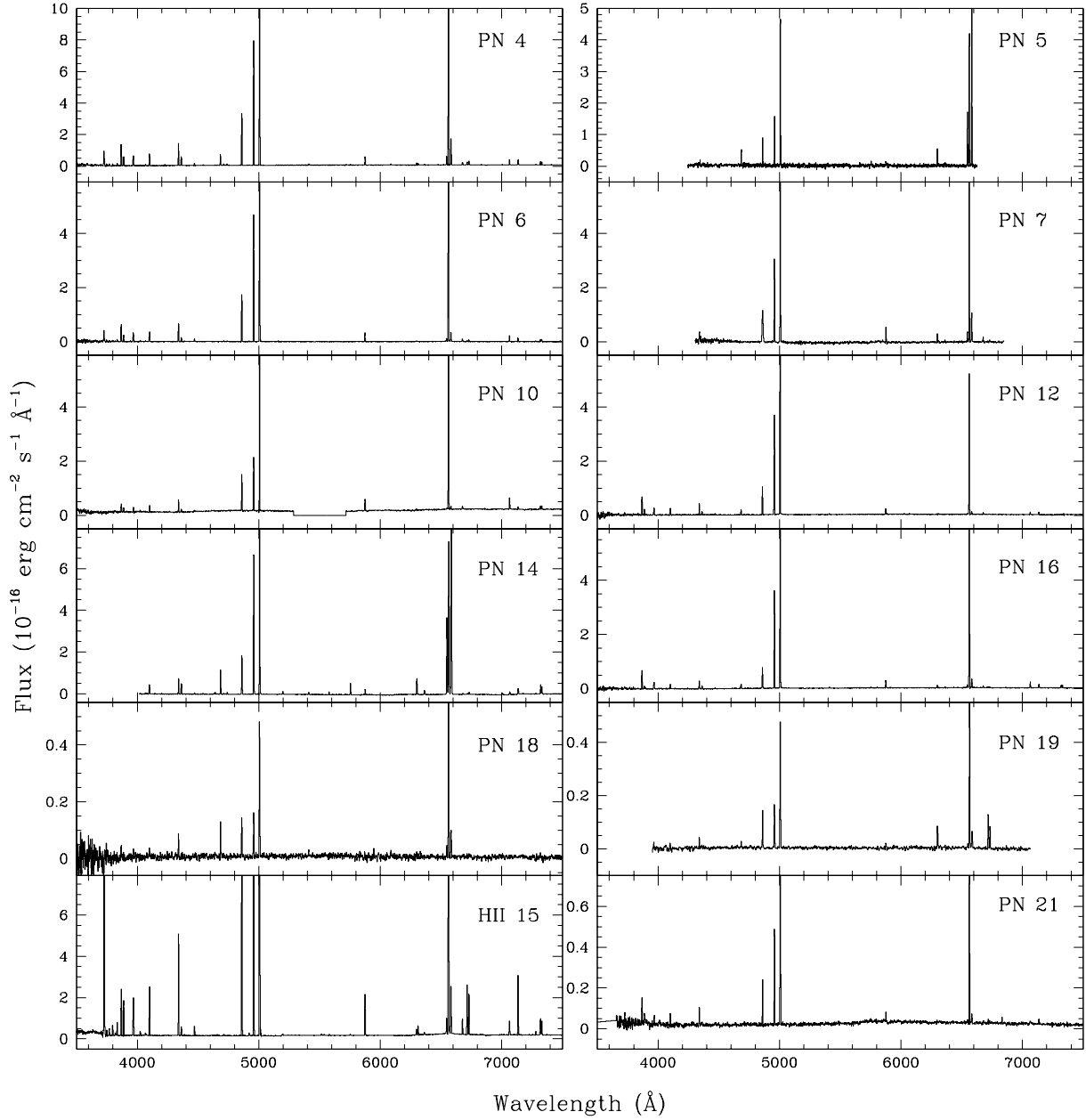}
\caption{Spectra of the PNe and the bright \ion{H}{ii}\,15.   Blue and red spectra have been combined.}
\end{center}
\end{figure*}

\section{Spectral Analysis}

  \subsection{Plasma diagnostic}
  The wide spectral coverage allowed us to determine electron temperature and densities from a couple of diagnostic sensitive ratios. In particular, electron temperatures were determined from the [\ion{O}{iii}]\,$\lambda\lambda$4363/5007 and [\ion{N}{ii}]\,$\lambda\lambda$5755/6583 ratios, when available. Only a few objects show the auroral [\ion{N}{ii}]\,$\lambda$5755  line intense enough to be measured, but most of the high-excitation nebulae present [\ion{O}{iii}]\,$\lambda$4363 with good signal-to-noise. Nebular densities were obtained through the [\ion{S}{ii}]\,$\lambda\lambda$6717/6731 line ratios and occasionally from [\ion{Ar}{iv}]\,$\lambda\lambda$4711/4740, although the latter one is very uncertain due to the faintness of lines. For these calculations, the IRAF routine `temden' of the package `nebular' (Shaw \& Dufour 1995) was used. The results are presented in Table 3.

 \subsection{Ionic and total abundances}
  To  derive ionic abundances, data were treated as homogeneously as possible. Ionic abundances of heavy elements, relative to H$^+$, were computed from the line intensities of Table 2 and the physical conditions described in the above section. In all cases we adopted an one-electron temperature scheme for all the ions by  using  T(\ion{O}{iii}) (the most confident electron temperature) as representative of the whole nebula. Only in the cases where  T(\ion{O}{iii}) could not be determined we used T(\ion{N}{ii}) if available. The electron density  derived from the [\ion{S}{ii}] sensitive ratio was used in all cases. The IRAF task `ionic«, from the package `nebular«, was used for these purposes.
  
 To determine He$^+$/H$^+$, we used three observed lines of He\,{\sc i} at $\lambda\lambda$4471, 5876 and 6678, weighted by 1:3:1. Case B He\,{\sc i} emissivities were taken from the collision-less (low-density limit) calculations by Bauman et al. (2005). We used a J-resolved code available on-line,\footnote{Available at http://www.pa.uKy.edu/~rporter//j-resolved/}  for these calculations. The collisional to recombination contribution was estimated from Kingdon \& Ferland (1995), using the interpolation formula provided by Porter et al. (2007). The effective recombination  coefficients for H$^+$ were taken from  Storey \& Hummer (1995). The He$^{++}$ abundance  was determined from the \ion{He}{ii} $\lambda$4686 recombination line, using the recombination coefficients by Storey \& Hummer (1995).
  
 The derived ionic abundances and their uncertainties are listed in Table 3. The uncertainties for heavy element ionic abundances are mainly due to uncertainties in the electron temperature, while for recombination lines, it is due to line intensities errors.
 
  To derive total abundances the unseen ions should be taken into account. For this we have adopted the ionization correction factors (icf) proposed by Kingsburgh \& Barlow (1994). Then total abundances correspond to the addition of observed ionic abundances multiplied by the corresponding icf. The final results are presented in Table 4, where we have separated PNe from \ion{H}{ii}  regions and we have included some results for well determined \ion{H}{ii}  regions from the literature.
  
\subsection{Supernova remnants?}
A few faint nebulae were marked by Leisy et al. (2005) as possible supernova remnants (SNR) due to their intense [\ion{S}{ii}] lines. Those were (in parenthesis Leisy et al. nomenclature) \ion{H}{ii}\,2 (\ion{H}{ii}\,21), \ion{H}{ii}\,4 (\ion{H}{ii}\,20), \ion{H}{ii}\,6 (\ion{H}{ii} 23) and \ion{H}{ii}\,9 (\ion{H}{ii}\,24). We obtained spectra of three of them (\ion{H}{ii}\,2, 4, and 9) finding they are very low excitation \ion{H}{ii}  regions, with very faint or none [\ion{O}{iii}] $\lambda$5007 emission and intense low excitation lines. The ionizing stars should be later than B0 stars. All these objects are of low density. We estimated the chemical composition for \ion{H}{ii}\,9 and\ion{H}{ii}\,4  by assuming  electron temperatures of 12000 K  and we found a normal chemical composition.

\setcounter{table}{2}
\begin{table*}
\begin{center}
{\footnotesize
\caption{Physical conditions and ionic abundances in PNe and \ion{H}{ii}  regions}
\begin{tabular}{lrrrrrrrrrr}
\hline
\hline
 &PN\,4~~ &PN\,5$^{1,2}~~$ &PN\,6~~ &PN\,7$^1$~~ &PN\,10~~ &PN\,12~~&PN\,14~~\\
\hline
T(\ion{O}{iii}) (K)&17870$\pm$980&$<$ 23000&12590$\pm$1500&$<$ 13400&17000$\pm$1000&13000$\pm$1000&18070$\pm$760\\
T(\ion{N}{ii}) (K)&15830$\pm$2300&16100$\pm$1700 &12082$\pm$1800 &&&&17550$\pm$1000\\
N(\ion{S}{ii}) (cm$^{-3}$)&2980$\pm$360&880$\pm$200&3910$\pm$480&4000$\pm$2000&1000 (adopt)&2000$\pm$1000&5150$\pm$500\\
N(\ion{Ar}{iv}) (cm$^{-3}$)&3240$\pm$1000&&1000:&&&2700$\pm$1000&15700$\pm$3500\\
\hline
He$^+$ &7.05$\pm$0.40 E-2&5.2$\pm$1.1 E-2&9.52$\pm$0.50 E-2&1.07$\pm$0.25 E-1&1.11$\pm$0.08 E-1&9.34$\pm$0.6 E-2&5.69$\pm$0.33 E-2\\
He$^{++}$(4686)&1.90$\pm$0.18 E-2&4.20$\pm$0.40 E-2&8.1$\pm$4.0 E-4&0.00 E+00&0.00 E+00&2.00$\pm$0.30 E-2&5.20$\pm$0.10 E-2\\
O$^+$(3727)&4.01$\pm$0.60 E-6&7.7$\pm$2.3 E-6$^3$&1.17$\pm$0.26 E-5&5.50$\pm$1.50 E-5$^3$&5.85$\pm$1.17 E-6$^3$&3.16$\pm$1.05 E-6&1.15$\pm$0.40 E-5$^3$\\
O$^{++}$(5007)&4.69$\pm$0.53 E-5&4.6$\pm$1.0 E-5&1.30$\pm$0.20 E-4&1.02$\pm$0.20 E-4&3.43$\pm$0.52 E-5&1.56$\pm$0.40 E-4&7.32$\pm$0.60 E-5\\
N$^+$(6583)&2.03$\pm$0.20 E-6&2.20$\pm$0.50 E-5&1.87$\pm$0.22 E-6&2.9$\pm$1.0 E-6&2.54$\pm$0.27 E-7&6.5$\pm$1.0 E-7&2.44$\pm$0.30 E-5\\
Ne$^{++}$(3869)&8.70$\pm$1.00 E-6&---&2.18$\pm$0.35 E-5&---&5.42$\pm$0.74 E-6&2.36$\pm$0.66 E-5&5.80$\pm$0.50 E-6$^5$\\
Ne$^{+4}$(3425)&2.92$\pm$0.45 E-8&---&0.00 E+00&---&0.00 E+00& --- & ---\\
S$^+$(6717)&1.03$\pm$0.10 E-7&5.5$\pm$1.3 E-7&1.14$\pm$0.17 E-7&7.8$\pm$2.0 E-8&---&2.54$\pm$0.60 E-8&7.93$\pm$1.00 E-8\\
S$^{++}$(6311)&8.43$\pm$1.20 E-7&2.3$\pm$1.0 E-6&8.45$\pm$1.65 E-7&7.75$\pm$2.00 E-7&---&1.54$\pm$0.50 E-6&4.1$\pm$2.0 E-7\\
Ar$^{++}$(7135)&1.96$\pm$0.17 E-7&4.10$\pm$0.80 E-7&3.12$\pm$0.32 E-7&4.9$\pm$2.0 E-7&7.9$\pm$1.4 E-8&2.82$\pm$0.44 E-7&2.38$\pm$0.30 E-7\\
Ar$^{+3}$(4740)&1.37$\pm$0.15 E-7&---&7.80$\pm$1.00 E-8&---&---&3.75$\pm$0.80 E-7&2.08$\pm$0.50 E-7\\
Ar$^{+4}$(7006)&3.21$\pm$0.40 E-8&---&---&---&---&---&1.34$\pm$0.20 E-7\\
\hline \hline
\multicolumn{5}{l}{$^1$ Observed with Gemini.}\\
\multicolumn{5}{l}{$^2$ T(\ion{N}{ii}) for ionic abundances.}\\
\multicolumn{5}{l}{$^3$ [\ion{O}{ii}] 7320 for O$^+$ abundance.}\\
\multicolumn{5}{l}{$^4$ Line intensity from Richer \& McCall (2007).}\\
\end{tabular}
}
\label{ionic1}
\end{center}
\end{table*}
\setcounter{table}{2}
\begin{table*}
\begin{center}
{\footnotesize
\caption{(cont.) Physical conditions and ionic abundances in PNe and \ion{H}{ii}  regions}
\begin{tabular}{lcrrrrrrrrr}
\hline
\hline
&PN\,16 &PN\,18 &PN\,19 &PN\,21 &\ion{H}{ii}\,15&\ion{H}{ii}\,9 &\ion{H}{ii}\,4$^1$\\
\hline 
T(\ion{O}{iii}) (K)&14390$\pm$720&18000$\pm$1000&20000$\pm$2000&13668$\pm$1500&11700$\pm$300&12000 (adopt)&12000 (adopt)&\\
T(\ion{N}{ii}) (K)&&&&&12400$\pm$1500\\
N(\ion{S}{ii}) (cm$^{-3}$)&1140$\pm$200&1420$\pm$400& $<$  100&184$\pm$100&187$\pm$100&$<$  100\\
N(\ion{Ar}{iv}) (cm$^{-3}$)&5100$\pm$2000\\
\hline 
He$^+$ &1.23$\pm$0.10 E-1&$<$ 6.00 E-2&1.30$\pm$0.50 E-1&1.00$\pm$0.10 E-1&8.23$\pm$0.05 E-2&4.5$\pm$1.4 E-2&1.50$\pm$0.50 E-2\\
He$^{++}$(4686)&2.50$\pm$0.25 E-2&6.5$\pm$1.6 E-2&1.50$\pm$0.30 E-2&0.00 E+00&0.00 E+00&0.00 E+00&0.00 E+00\\
O$^+$(3727)&1.73$\pm$0.50 E-5$^3$&--- &1.64$\pm$0.40 E-5$^3$&5.5$\pm$1.5 E-6&4.31$\pm$0.30 E-5&---&---\\
O$^{++}$(5007)&1.39$\pm$0.20 E-4&1.88$\pm$0.30 E-5&1.82$\pm$0.40 E-5&8.3$\pm$2.0 E-5&8.36$\pm$0.40 E-5&---&---\\
N$^+$(6583)&1.40$\pm$0.20 E-6&2.13$\pm$0.40 E-6&1.17$\pm$0.50 E-6&1.07$\pm$0.30 E-6&1.23$\pm$0.20 E-6&---&---\\
Ne$^{++}$(3869)&2.56$\pm$0.30 E-5&4.01$\pm$0.70 E-6&---&1.51$\pm$0.40 E-5&1.54$\pm$0.20 E-5&---&---\\
%Ne$^{+4}$(3425)\\
S$^+$(6717)&4.42$\pm$0.60 E-8&1.58$\pm$0.15 E-7&6.7$\pm$1.1 E-7&1.44$\pm$0.40 E-7&2.72$\pm$0.30 E-7&---&---\\
S$^{++}$(6311)&6.26$\pm$1.00 E-7&---&8.8$\pm$1.2 E-7&$<$ 3.00 E-6&2.47$\pm$0.30 E-6&---&---\\
Ar$^{++}$(7135)&2.57$\pm$0.30 E-7&---&2.26$\pm$0.60 E-7&3.37$\pm$0.50 E-7&6.23$\pm$0.15 E-7&---&---\\
Ar$^{+3}$(4740)&3.99$\pm$0.40 E-7 &---&---&---&---&---&---\\
%Ar$^{+4}$(7006)\\
\hline \hline
%\multicolumn{5}{l}{$^1$ Observed with Gemini.}\\
%\multicolumn{5}{l}{$^2$ T(NII) for ionic abundances.}\\
%\multicolumn{5}{l}{$^3$ [\ion{O}{ii}] 7320 for O$^+$ abundance. }\\
\end{tabular}
}
\label{ionic2}
\end{center}
\end{table*}

\begin{table*}
\begin{center}
{\footnotesize
\caption{Total abundances for PNe and \ion{H}{ii}  regions in NGC\,6822  and other galaxies$^1$}
\begin{tabular}{lllllllll}
\hline
\hline
object&  ~~ He/H  &O/H&~~N/H& ~~Ne/H& ~~S/H& ~~Ar/H& other IDs$^2$ , comments$^3$\\
\hline 

PN\,12& 0.113$\pm$0.009 &8.26$\pm$0.10 &7.57$\pm$0.13 &7.44$\pm$0.10 &6.62$\pm$0.15 &5.83$\pm$0.10 & PN14-S14, young\\
PN\,16& 0.148$\pm$0.010 &8.25$\pm$0.08 &7.16$\pm$0.20 &7.51$\pm$0.10 &6.02$\pm$0.20 &5.82$\pm$0.12 & PN13-S16, young\\
PN\,7& 0.107$\pm$0.015 &8.20$\pm$0.15 &6.92$\pm$0.17 &---&5.98$\pm$0.20 &5.96$\pm$0.13 & PN2, young\\
PN\,6& 0.096$\pm$0.005 &8.15$\pm$0.07 &7.36$\pm$0.15 &7.38$\pm$0.12 &6.20$\pm$0.29 &5.63$\pm$0.10 & PN1, old\\
PN\,14& 0.109$\pm$0.006 &8.12$\pm$0.08 &8.44$\pm$0.14 & --- &5.89$\pm$0.30 &5.81$\pm$0.12 &  PN7-S33, Type I, young\\
PN\,21& 0.106$\pm$0.012 &7.95$\pm$0.10 &7.24$\pm$0.15 &7.21$\pm$0.10 &$<$ 6.75&5.80$\pm$0.12 &  PN12, young\\ 
PN\,5& 0.094$\pm$0.015 &7.90$\pm$0.10 &8.36$\pm$0.20 &---&6.64$\pm$0.29 &5.88$\pm$0.15 &  PN3, Type I, young\\
PN\,4& 0.090$\pm$0.005 &7.78$\pm$0.07 &7.48$\pm$0.12 &7.05$\pm$0.10 &6.22$\pm$0.15 &5.59$\pm$0.10 &  PN4,  Type I?, old\\
PN\,10& 0.111$\pm$0.010 &7.60$\pm$0.08 &6.30$\pm$0.13 &6.80$\pm$0.10 &---&5.20$\pm$0.15 &  PN19, old\\
PN\,19&0.145$\pm$0.040 &7.57$\pm$0.12 &6.42$\pm$0.15 &---&6.22$\pm$0.20 &5.63$\pm$0.15 &  PN9, old\\
PN\,18&$<$ 0.124 &7.49$\pm$0.15 &---&6.82$\pm$0.15 &---&---&  PN11, old\\
\hline
$<$non-type I$>$& 0.114$\pm$0.014  & 7.92$\pm$0.18  & 7.06$\pm$ 0.31  & 7.17$\pm$0.20  & 6.21$\pm$0.17  & 5.68$\pm$0.15 & Type I PNe excluded, 2$\sigma$ err\\
\hline
\ion{H}{ii}\,15 & 0.090$\pm$0.005 &8.10$\pm$0.05 &6.56$\pm$0.08 &7.36$\pm$0.07 &6.49$\pm$0.10 &5.97$\pm$0.08 & \ion{H}{ii}8-S28\\
H\,V& 0.084$\pm$0.002 &8.08$\pm$0.03 &6.85$\pm$0.15  &7.35$\pm$0.03 &6.61$\pm$0.05 & 5.84$\pm$0.03 &   PPR\\
H\,X& 0.084$\pm$0.002 &8.01$\pm$0.05 &6.76$\pm$0.16 &7.27$\pm$0.05  &6.51$\pm$0.06  &5.84$\pm$0.05 &  PPR\\
H\,I & --- &		7.98$\pm$0.09 & ---	& 7.34$\pm$0.19 &---&--- &		 LSV \\	
K$\alpha$ & --- & 8.00$\pm$0.04 & 6.16$\pm$0.10 & 7.51$\pm$0.10 &---  & 6.04$\pm$0.07& LSV\\
K$\beta$ & --- & 8.18$\pm$0.07 & ---& 7.60$\pm$0.13 & ---&--- &  LSV \\
KD28e	&--- &	8.19$\pm$0.05 & 6.38$\pm$0.06  &	7.64$\pm$0.09 & ---& 6.10$\pm$0.05 &	 	 LSV\\
\hline
$<$\ion{H}{ii} regions$>$&0.086$\pm$0.004  &8.06$\pm$0.04  &6.72$\pm$0.14  & 7.33$\pm$0.05  &6.54$\pm$0.06  & 5.88$\pm$0.07 & 3 best observed obj., 2$\sigma$ err \\	
\hline \hline
LMC-\ion{H}{ii} & 0.085 &8.40 & 6.90  &7.60 & 6.70 & 6.20 & D89, G99\\ 
LMC-PNe& 0.091&8.33  &7.45 &7.54 & 6.93 &5.93& non-type I, LD06\\
SMC-\ion{H}{ii} & 0.079  & 8.00  & 6.50 & 7.20  & 6.30& 5.90&  D89, G99 \\
SMC-PNe&0.079 & 8.09  &7.11 &7.16 & 6.62  & 5.51 & non-type I, LD06\\
%SMC-PNe&0.093 & 7.92 & 7.38 & 7.17 & 6.08 & 5.46 & PSR\\
NGC3109-\ion{H}{ii} & 0.085 & 7.77  &6.44 &6.87 &6.39 & 5.83 & PSR\\
NGC3109-PNe  & 0.087 & 8.16  & 7.49 & 7.18  & 6.26  & 5.66 & PSR\\
%Orion&&8.60&&&&& Esteban 1998\\
Sun&0.085&8.66& 7.78 & 7.84& 7.14& 6.18& Grevese  et al. (2007)\\
\hline \hline
\multicolumn{8}{l}{$^1$ Heavy elements in 12\,+\,log\,(X/H), objects ordered by 12\,+\,log\,(O/H) values.}\\
\multicolumn{8}{l}{$^2$ IDs: PN\# from Leisy et al. (2005) and S\# from Killen \& Dufour (1982)}\\
\multicolumn{8}{l}{$^3$ Ref: PPR: Peimbert et al. (2005); LSV: Lee et al. (2006),  D89: Dennefeld (1989); G99: Garnett (1999);}\\
\multicolumn{8}{l}{\hskip 0.85cm LD06: Leisy \& Dennefeld (2006); PSR: Pe\~na et al. (2007).}\\
\end{tabular}
}
\label{tab:abundances}
\end{center}
\end{table*}
 
% For the galaxy: Wang \& Liu (MNRAS, Feb 2008) Ne/O (disk PNe)= 0.24, Ne/O (bulge PNe)=0.25, Ne/O (Orion)=0.24 (t$^2$=0.00)

\section{Chemical abundances}
\subsection{Comparison with previous determinations}

From our data we have determined He, N, O, Ne, S and Ar for most of the observed objects. Our sample contains 4 PNe  in common with Richer \& McCall (2007, hereafter RMcC), which are   PN\,10 (their PN\,19), PN\,21 (their PN\,12), PN\,14 (their S\,33) and PN\,16 (their  S\,16). We observed and re-observed as many PNe as possible in  order to secure a homogeneous and more precise data.  Comparing our derived  physical conditions and chemical abundances with those by RMcC, we found that:

$\bullet$ The physical conditions, particularly the electron temperature, are very similar. Densities  are similar within uncertainties.

$\bullet$ Our He abundances are equal within uncertainties for all the objects but PN\,16, where they found a He/H ratio 40\% lower. Possibly this is due to their much higher reddening that largely affects the \ion{He}{i} $\lambda$5876 intensity.

$\bullet$ The O abundances are very similar. Their O/H values differ from ours only by 0.06 dex at the most. But other elements are not so consistent.
 Their N/O for PN\,14  is too large (7.94) and for PN\,21 (0.03) is too low, compared to our values. These differences are mainly due to N/O abundance ratio is based on N$^+$/O$^+$ ionic abundance ratio, being both ions very under-abundant in high excitation PNe. In particular the O$^+$ abundance is very uncertain due to uncertainties in the [\ion{O}{ii}] $\lambda$3727 line which in general is weak, it depends strongly on the adopted temperature and density, and it is very affected by errors in the adopted reddening.  About Ne/O ratios, the differences are  small except for one object: PN\,21,  for which we obtain Ne/O=0.18 and RMcC derived 0.363 with a huge error. We consider our determination more confident due to the higher S/N of our [\ion{Ne}{iii}] lines.

\subsection{ Abundances in \ion{H}{ii}  regions: Is the ISM in NGC\,6822 chemically homogeneous?}

As said in the introduction, no conclusive evidence  for  chemical inhomogeneities in the present ISM of NGC\,6822 has been found (Lee et al. 2006). To  further analyze this, we compare  our results for  \ion{H}{ii}\,15 with  the other two best determined results for \ion{H}{ii}  regions from the literature which are H\,V, and H\,X. Their abundances were taken from collisionally excited lines determinations by Peimbert et al. 2005). The three regions show, in average $<$12\,+\,log(O/H)$>$ = 8.06$\pm$0.04. If we include the values reported by Lee et al. (2006) for their objects with [\ion{O}{iii}] $\lambda$4363 measured (presented in Table 4), we still found a very chemically homogeneous galaxy, with an average abundance of $<$12\,+\,log(O/H)=8.08$\pm$0.06$>$. All these \ion{H}{ii}  regions belong to the optical zone of NGC\,6822, with \ion{H}{ii}\,15 very near (de-projected distance 62 pc) to the galactic center proposed by Brandenburg  \& Skillman (1998) while H\,V and H\,X are much to the North at de-projected distances of about 1.01 and 1.08 kpc from the center (half-way to the border of the H\,{\sc i} disk), H\,1 is 1.78 kpc to the North-West  and K$\alpha$ is at 1.30 kpc to the West.  Thus, our conclusion is that the present ISM is chemically homogeneous, at least in the  2 kpc around  the center. Other heavy elements like N, Ne, S and Ar are also very homogeneous for the three best observed \ion{H}{ii} regions, showing averages values of $<$N/O$>$= 0.05$\pm$0.02, $<$Ne/O$>$= 0.184$\pm$0.005, $<$S/O$>$=0.029$\pm$0.005 and $<$Ar/O$>$=0.007$\pm$0.002. The present ISM chemical abundances  in NGC\,6822 (based on the  three best observed \ion{H}{ii} regions) are closer to those in the SMC than to those in the LMC (see Table 4), except for N/H which lies  just in the middle. The N/O abundance ratio of 0.005 is larger than the 0.003 value of both Clouds. Thus, NGC\,6822 could be a more evolved galaxy than the Magellanic Clouds (similar results were obtained Peimbert et al. 2005 from their C/O ratio).

\subsection{PN abundances}
For our PN sample, the abundances present  a different situation than \ion{H}{ii} regions. In Fig. 2 we present the behavior of 12\,+\,log\,(Ne/H) vs. 12\,+\,log\,(O/H), and 12\,+\,log\,(Ar/H) vs. 12\,+\,log\,(O/H) abundances for all our PNe. The average values for  \ion{H}{ii} regions are marked. 

 In the upper panel of this Figure it is observed that the usual strong correlation between Ne and O abundances is very well defined for PNe and  \ion{H}{ii} regions in NGC\,6822. The slope of our linear fit is very near 1 and it is equal to the slopes reported for other galaxies; e.g., in the Milky Way a slope near 1 was reported  by Henry (1989);  Stasi\'nska et al. (1998) report a very tight Ne-O correlation (with slope of about  1) for a PNe sample  of several galaxies; the PNe in the LMC and SMC show slopes of 1.13 and 1.08 respectively (Leisy \& Dennefeld 2006), while in M33 Magrini et al. (2009) report a slope of 0.90 for their PN sample.  The average value $<$Ne/O$>$, however, seems to vary with metallicity, being larger in richer galaxies (Wang \& Liu 2008).

 Also Fig. 2 shows that O/H values in PNe are dispersed in a large range, from low-metallicity objects with O/H abundances 0.5 dex poorer than the average of \ion{H}{ii}  regions   up to 0.2 dex richer than this average. PNe are faint objects so their uncertainties in abundances are larger than in the \ion{H}{ii}  regions analyzed, however  the dispersion is real, considering the uncertainties in abundances determinations. The richer objects, with 12\,+\,log\,(O/H) in the range 7.9--8.2, could be considered as having abundances similar  to \ion{H}{ii}  region abundances, but the objects in the low metallicity limit (12\,+\,log\,(O/H) $<$\,7.8) belong to an older less-enriched population. 

 As it was said in the introduction,  O and Ne original abundances in PNe can be perturbed, either by O destruction  through ON-cycle  in more massive progenitors, or when freshly synthesized O (and Ne) is  dredged up to the atmosphere of low metallicity PN progenitors (third dredge-up). The limit in metallicity for such an enrichment is still debated; thus, so far for NGC\,6822, neither of these elements can be considered safe for testing the original chemical composition of PN progenitor stars and the use of other elements, like S and Ar, is necessary. 

For most of our objects, S and Ar were determined, although these elements have large uncertainties.   In particular, S abundance  has large uncertainties due to only S$^+$ and S$^{++}$  are observed, being  these ions  under-abundant in high excitation PNe. Besides, the icf's proposed by Kingsburgh \& Barlow (1994) depends on O$^+$, also an under-abundant ion. Therefore, S abundances are not very reliable.  
On the other hand, Ar is better determined as the observed ions (Ar$^{++}$ and Ar$^{+3}$) are the most abundant species  in high excitation PNe and the ifc's are small. In Fig. 2 (lower panel) Ar vs. O  abundances are presented. Contrary to what happens with Ne, there is no obvious Ar to O relation, except for an increasing trend which is expected and that also appears in the PN samples of the other galaxies mentioned. The important fact here is that no PN presents an Ar/H abundance ratio larger than the average value of  \ion{H}{ii} regions. Besides, PNe can be divided roughly  in two groups: one with Ar/H ratio similar to  \ion{H}{ii} regions (all these objects also present large O/H ratios) and other with  low Ar/H values. In the following we will discuss PN abundances in terms of these two groups.
\begin{figure}%[ht]
\begin{center}
\label{}
\includegraphics[width=10cm,height=8cm]{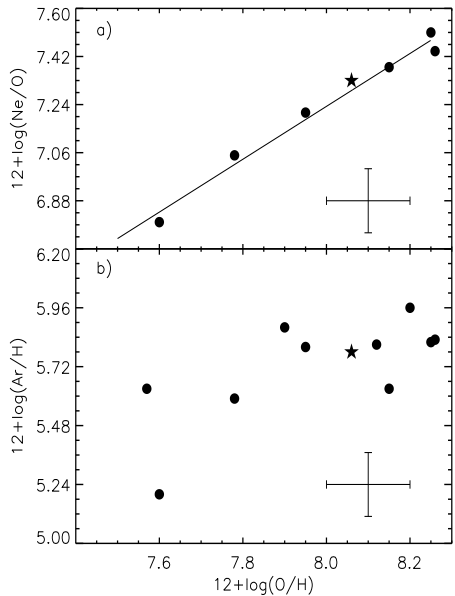}
\caption{ The behavior of 12\,+\,log\,(Ne/H) and 12\,+\,log\,(Ar/H) vs. 12\,+\,log\,(O/H) is presented. Filled dots are PNe and the star is the average value for \ion{H}{ii} regions. The tight linear correlation between Ne and O corresponds to 12\,+\,log(Ne/H)=0.99\,(12\,+\,log(O/H))$-$0.72, with a correlation coefficient R$^2$=0.98 (after removing the more deviant point PN\,18). Ar/H determination is much more uncertain and the correlation is not as clear. But it is found that not PNe has Ar/H larger than \ion{H}{ii} region value, which is not the case for Ne/H  and O/H.}
\end{center}
\end{figure}

\subsection{The N abundances and Type I PNe} 

In Table 4 we find that all PNe, but two, show N-enrichment relative to the average value for \ion{H}{ii}  regions. 
Similar results are found for all galaxies and this is usually interpreted as result of the different dredge-up episodes 
in the PN progenitor star, where some products from nucleosynthesis (mainly He, C and N) are brought to the surface. The N-enrichment in PNe indicates 
that these objects are a major source of N production in a galaxy, but for NGC\,6822, such nitrogen has not 
enriched the ISM  yet.

  N-enrichment occurs in the stellar surface during the AGB dredge-up events 
 and it has been attributed to CN-cycle where original C is transformed into N (for the more massive progenitors, also O can be transformed to N through the ON-cycle, e.g. Henry 1990, and nebulae show a depleted O). In addition, very N-rich nebulae (usually named Type I) can be produced by the more massive 
progenitors that can experience envelope-burning conversion to nitrogen of 
dreged-up primary carbon. Different criteria has been proposed to define these particularly N-enriched PNe, 
Peimbert  (1978) classified as  Type I PNe  those galactic nebulae with 
He/H\,$>$\,0.125 and N/O$>$0.5, while Kingsburgh \& Barlow (1994) defined Type I PNe as those nebulae 
having N abundances that exceed the total C+N abundance of \ion{H}{ii} regions in the same galaxy 
and suggested a limit of  N/O\,$>$\,0.8 for our galaxy, independently of the He abundance. 
These criteria would required some corrections for PNe in low-metallicity galaxies, e.g., 
 Leisy \& Dennefeld (1996) suggested a limit N/O\,$>$\,0.25 for  Type I PNe in
the Magellanic Clouds.  In our case we will employ Kingsburgh \& Barlow criteria, by considering that Peimbert et al. (2005) 
found log C/O\,=\,$-0.31\pm0.13$ for H\,V in NGC\,6822  and  that N/O\,=\,0.05 in \ion{H}{ii} regions. This would imply that a   Type I PN in this galaxy  should have N/O\,$\geq$\,0.49
(there is a large margin of uncertainty in this value as, considering Peimbert et al. 2005 uncertainties for C/O, the limit of N/O for a Type I PN is in the range from 0.36 to 0.66). 
Thus,  in our sample we find clearly two extreme Type I PNe (PN\,5 and PN\,14) with N/O\,$>$\,2, and a possible third Type I PN, PN\,4, with N/O\,$\sim$\,0.50. 

The O and Ar abundances in the extreme Type I PNe, PN\,14 and PN\,5, are equal, within uncertainties, 
to the average values of \ion{H}{ii}  regions, thus confirming that they are young objects. 
The third Type I candidate, PN\,4, which is the brightest PN observed in this galaxy,  requires a closer look. All its elemental abundances, but N, 
are about a factor of 2 lower than the averages in \ion{H}{ii}  regions, therefore this object
should have emerged in a less enriched medium, a few Gyrs ago and probably its progenitor 
is not a very massive one. Besides it should be taken into account that the N/O abundance ratio
 is determined via the N$^+$/O$^+$ ionic abundance ratio which is very temperature dependent. 
 For PN\,4 we have determined both, T(\ion{O}{iii})  and T(\ion{N}{ii}), electron temperatures 
 (although T(\ion{N}{ii}) has a large uncertainty) then, to test our abundance ratio, 
we have calculated  N$^+$ and O$^+$ ionic abundances  using T(\ion{N}{ii}). We found a N$^+$/O$^+$  abundance 
ratio of 0.45 instead of 0.50. Then, according to our criteria, PN\,4 is in the limit and can be considered only marginally as a Type I PN.

\subsection{Non Type I PNe}
Among the non Type I nebulae (those that are not N-rich), there are a few (PN\,7, PN\,12,  and PN\,16) showing O/H abundance ratios about 0.15-0.20 dex 
higher than \ion{H}{ii}  regions. Also the Ne/H values in PN\,12 and PN\,16 seem to be slightly enhanced, relative to \ion{H}{ii} regions. On the other hand,
 the three nebulae present   Ar abundances equal, within uncertainties, to \ion{H}{ii} regions (see Table 4 and  Fig. 2). Then these PNe could be   O and Ne-enriched nebulae, via the third dredge-up. However the effect is small and other explanations are possible, for example, O locked in dust grains in \ion{H}{ii} regions. Based on their Ar abundance, we will  consider these nebulae as belonging to the young PNe. 

Three non-type I objects, PN\,10, PN\,18 and PN\,19, are O, Ne and Ar poorer than \ion{H}{ii}  regions by factors larger than 2. Together with the  Type I candidate, PN\,4, these objects correspond to bona-fide low-metallicity PN progenitors and they belong to an older population formed in a less enriched medium. In this group we will also include PN\,6, which has an  O/H similar to \ion{H}{ii} regions but a much lower Ar abundance. 

\subsection{Comparison with  samples in other galaxies}
 In Table 4 we present the average abundances values for the three best determined \ion{H}{ii} regions and the whole sample of not-type I PNe. For comparison,  we have included the respective values for the the Magellanic Clouds as derived by Leisy \& Dennefeld (1996, 2006) and also the abundances derived for the low metallicity irregular  galaxy NGC\,3109 and the solar  values. In general, the chemical abundance pattern of PNe in NGC\,6822 follows  closely the behavior of PNe in the Magellanic Clouds. The average abundance values of heavy  elements in non-type I PNe  in NGC\,6822 are very similar to the ones in the SMC. In this case (and contrary to \ion{H}{ii} regions) also the N/H ratio shows similarity with the SMC, indicating similar nebular enrichment  due to stellar nucleosynthesis. Differently than PNe in NGC\,3109, the PN sample in NGC\,6822 do not  show O-enrichment relative to \ion{H}{ii} regions. For NGC\,3109, this O-enrichment could be attributed to newly synthesized O dredged-up by convection in the progenitor envelopes. As already said, such an   enrichment happens mainly in low metallicity environments (12\,+\,log (O/H) = 7.77 in \ion{H}{ii} regions of NGC\,3109). Apparently the metallicity of NGC\,6822 is not low enough for detecting such an event.
 
 Non Type I PNe in NGC\,6822 appear richer in He than in the other galaxies but this could be an artifact due to a couple of apparently very He-rich nebulae: PN\,16 and PN\,19. The latter one has a large uncertainty in the He/H determination and if it is excluded from the average, we obtain a He/H average value of 0.096$\pm$0.006, still large but, within uncertainties it is similar to the value in the LMC. Interestingly,  contrary to expectations, the two extreme Type I PN are not particularly He-rich.

\section{Chemical enrichment in NGC\,6822. A preliminary evolution model.}

\subsection{Observational constraints}
In \S 4 we present the chemical abundances of several \ion{H}{ii}  regions taken from 
diverse authors, including our \ion{H}{ii}\,15 determinations, as derived from collisionally excited lines.
Those \ion{H}{ii}  regions are located in different parts of the optical area
and we have found no evidence of chemical inhomogeneities in this galaxy.
For that reason we choose the average $<$12\,+\,log\,(O/H)$>$ = 8.06$\pm$0.04 
 as representative
of the current O/H abundance ratio of the whole galaxy
and we have computed several chemical evolution models of NGC 6822
to reproduce this O/H value. Notice that no O locked in dust grains has been considered.

 Abundances of PNe in our sample will be used as additional constraints.
As discussed in \S 4, several PNe in our sample belong to  the young population in this galaxy and their average O/H ratio is  similar to the one of \ion{H}{ii} regions. Their other elements are also similar to \ion{H}{ii} regions, therefore this PN sample does not provide further constraints, but there is a sample of older  PNe, with much lower O, Ne and Ar  that will help to test the model predictions at the time of formation of their progenitors. The elements used for constraining the model should be those that clearly represent the initial  abundances in the progenitors. It is known that  nucleosynthesis and convective processes in LIMS  perturbed mainly He, C, and N abundances in PNe. We have not determined C abundances for our PNe, but we do derive He and N abundances which will not be use as observational constraints in our model. In addition, as explained in the introduction,   the O/H ratio could have been  affected by processes such as ON-cycle or third dredge-up, therefore it should be cautiously considered as a confident tracer of the original  O abundance. On the other hand, the production of  S, Cl, and  Ar by LIMS is negligibly, so these elements
 can  be considered as important observational constraints of the ISM in the past. For our PN sample 
we only have S and Ar abundances, but only Ar determinations are reliable. Thus, for our chemical 
evolution modelling  we will rest mainly on Ar abundances to support our conclusions 

Thus,  based mainly on their Ar abundances, 
we have divided the PNe of Table 4 in two groups: `young' and `old'.
PN\,5,  PN\,7, PN\,12, PN\,14, PN\,16 and PN\,21, 
which show $<$12\,+\,log\,(O/H)$>$=8.10$\pm$0.10 and $<$12\,+\,log\,(Ar/H)$>$=5.85$\pm$0.06, 
are young objects.  
PN\,4, PN\,6, PN\,10, PN\,18, and PN\,19, 
with $<$12\,+\,log\,(O/H)$>$=7.71$\pm$ 0.20 and $<$12\,+\,log\,(Ar/H)$>$=5.50$\pm$0.13, 
are PNe with old progenitors. 
Following  Allen et al. (1998), we will assume ages between 1 to 3 Gyr for the young PN population and  between 3 and 9 Gyr for the old one. The present age of the galaxy  is assumed to be 13.5 Gyr.

In next section we explain briefly the code we have developed to compute chemical evolution models. 
A preliminary model has been computed for NGC\,6822, which fits quite well  some of the constraints mentioned above. 
In a following paper (Hern\'andez-Mart\'inez et al., in preparation)  
we will discussed extensively other more detailed models with their assumptions and results.  

\subsection{Chemical evolution models}

\subsubsection{Antecedents }

Carigi et al. (2006, hereafter CCP) computed chemical evolution models for NGC\,6822 using
CHEVO code, which follows the lifetime of each star formed.
These models were built to reproduce the gaseous mass of  the galaxy and  the abundance value 
12\,+\,log\,(O/H)=8.42$\pm$0.06, as determined  by Peimbert et al. (2005) for Hubble V (H\,V), the brightest \ion{H}{ii}  region of the galaxy.
 Peimbert et al.  X$_{\rm i}$/H abundance ratios  are higher than ours, because their determinations are 
from recombination lines or collisionally excited lines considering a temperature fluctuations, $t^2 > 0$
 and dust correction.  
CCP also considered as constraints the C/O, N/O and Fe/O abundance ratio, derived also with $t^2 > 0$.

The code used by CCP allows the evolution of 6 elements only. Then they computed models for few chemical elements. Moreover, their model results could only be compared with observational data at present time 
because that work  was based on abundances  from \ion{H}{ii}  regions exclusively.
In this work, we are able to explore a more precise model due to 
we compute a model for several chemical elements,
as many observational data as stellar yields we get, and
we study the whole chemical evolution of NGC\,6822
comparing our model results with  
current and past observational data from \ion{H}{ii}  regions and PNe.

%INGREDIENTS
\subsubsection{Our code}

The code developed by one of us (L.H.-M.), and applied to NGC\,6822, takes into account the delayed contribution of 
all LIMS   as represented by one `average' star that enriches the interstellar medium,
while it is considered that all massive stars (MS)  enrich the 
ISM instantaneously, in a similar way as presented by Franco \& Carigi (2008).
The  new code was developed considering  a general star formation, 
initial mass function, infalls, outflows, and stellar yields for 27 elements.
Moreover this code includes the progenitors of Type Ia supernovae (SNIa),
main producers of Fe,  unlike models by Franco \& Carigi (2008).

Our model  assumes:

i) the star formation history (SFH) obtained by CCP,
that reproduces the photometric properties of NGC\,6822,

ii) the initial mass function by Kroupa, Tout \& Gilmore (1993, here after  KTG IMF) 
between 0.1 and 40 M$_\odot$, in order to reproduce the current 12\,+\,log\,(O/H)\,=\,8.06$\pm$0.04

iii) the infall computed by CCP predicted by a $\Lambda$CDM cosmology,

iv) a well-mixed outflow during 5.3 Gyr, in order to reproduce the present-day gas mass 
observed in NGC\,6822, (1.98 $\pm$ 0.02)$\times$10$^8$ M$_\odot$ (see CCP)
and 

v) a set of metallicity $Z$ dependent yields for 27 elements formed by massive star yields, 
low and intermediate mass star yields, and Type Ia SN yields.

For massive stars (M $>$ 11 M$_\odot$)
we consider yields by pre-SN stage and SN stage.
The pre-SN yields are taken from Geneva group work 
(Hirshi 2007; Meynet \& Maeder 2002, Hirshi et al. 2005). 
The SN yields are taken from Woosley \& Weaver (1995) adopting their
models B, for 12 to 30 M$_\odot$ and their models C, for 35 to 40 M$_\odot$.
We combine Geneva yields with Woosley \& Weaver  yields using the prescription
explained by Carigi \& Hernandez (2008).
The SN explosions do not contribute to He, C, N, and O yields,
but they do to heavier elements, like  Ne, S, Cl, Ar, and Fe. 

For LIMS ( 1 $<$ M $<$ 6 M$_\odot$) we adopted the
new yields by Karakas \& Lattanzio (2007). 
 
Their models were calculated from the zero-age main sequence to 
near the end of the thermally-pulsing asymptotic giant branch including
the PN phase.
Their code is quite complete and considers the core He-flash, the third dredge-up, and the hot bottom burning, 
during a total number of thermal pulses between 20 and 100 approximately,
 that increases with the initial mass
and decreases with the initial metallicity.
They assume a complex CNO treatment and their models predict O production only
for very metal-poor LIMS ($Z=1.0 \times 10^{-4} = 0.005 \,Z_\odot $).
For the rest of initial stellar metallicities their models predict O destruction,  
opposite to Marigo's model  (Marigo 2001).
They do not considered scaled-solar compositions and therefore their yields are more reliable
to study object with chemical abundances similar to the Magellanic Clouds, like NGC 6822.

Since the mentioned  yields for high and low mass stars do not cover the 6$-$11 M$_\odot$ range we
extrapolate the MS and LIMS yields to 7.5 M$_\odot$.

Iron abundance in a galaxy is mainly produced by Type Ia supenovae. A SNIa originates by a binary system of LIMS,
after the most massive member become C-O white dwarf
and accretes material from its companion in any evolutionary stage 
(main sequence, red giant, or C-O white dwarf).
But the exact physical mechanisms are not clear
and consequently the time delay of that type of SN are still unknown.

For SNIa, 
we assume that a fraction, $A_{bin}$, of the stars with masses between 3 and 15 M$_\odot$
give up binary systems and every one of those systems becomes a SNIa.
All Type Ia supernovae of each stellar generation
enrich the interstellar medium 1 Gyr after formation of SNIa progenitors,
according to Fig. 1 of Mannucci (2007).
Each SNIa,
independent of the initial metallicity, ejects
heavy elements, mainly Fe, with an efficiency (yield) given by the
model W7 from Nomoto et al. (1997).

The fraction $A_{bin}$ is a free parameter of the model 
and it is obtained in order to reproduce the average Fe observed in 2 stars of NGC\,6822
($<$[Fe/H]$>\,=-$0.49$\pm$0.22 in solar units, Venn et al. 2001).
Based on all previous assumptions, we get $A_{bin}=0.01$, that is, 
the 1 \% of 3-15 M$_\odot$ stars becomes SNIa.

%RESULTS. FIG 3
\begin{figure*}%[ht]
\begin{center}
\label{cem}
\includegraphics[width=15cm,height=11cm]{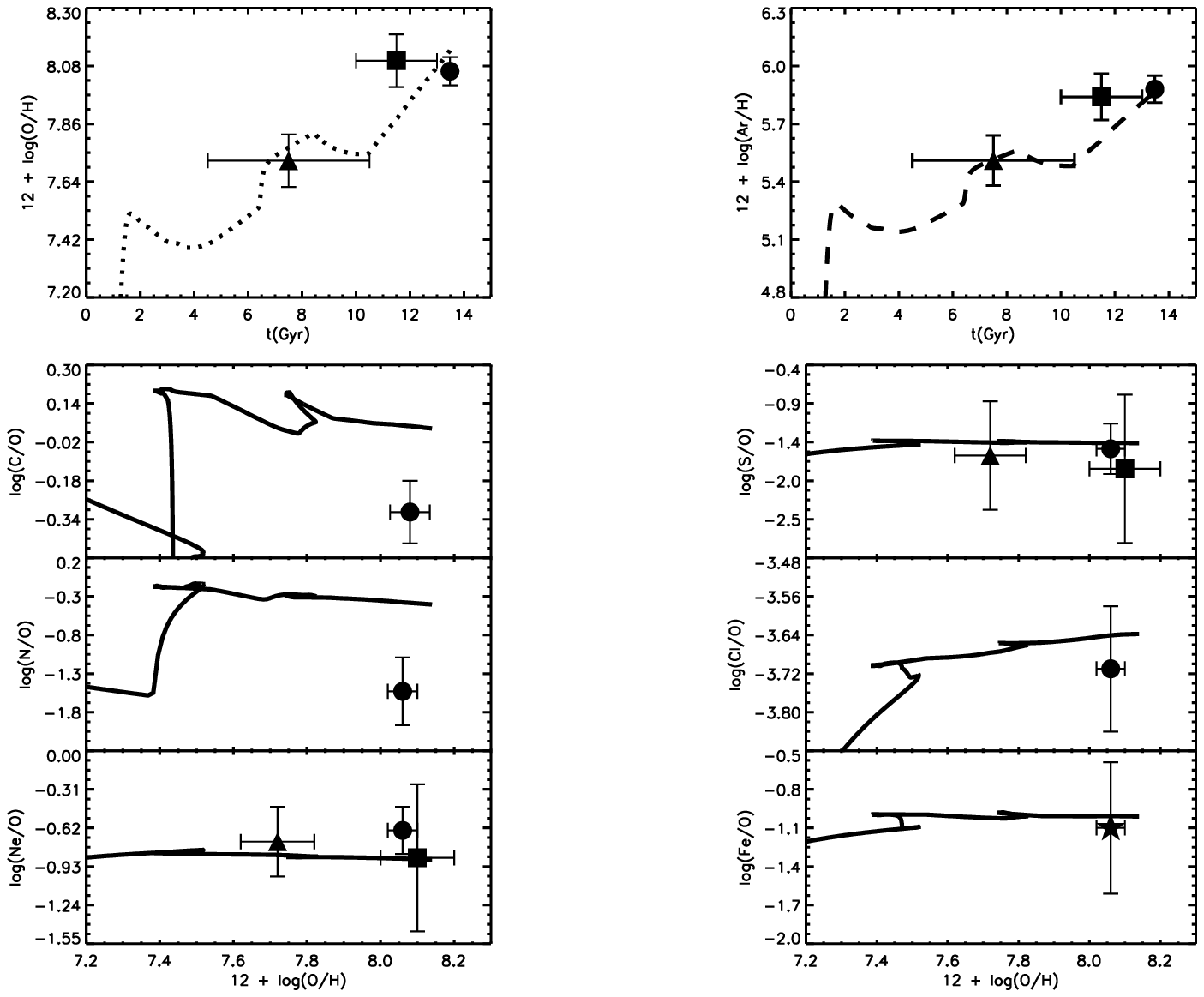}
\caption{ Evolution of 12\,+\,log\,(O/H) and 12\,+\,log\,(Ar/H) vs. time
 and C/O, N/O, Ne/O, S/O, Cl/O, Ar/O and Fe/O vs. 12\,+\,log\,(O/H) as predicted by 
one of our best chemical evolution models for NGC 6822 (discussed in text). The observational data show the average abundance values  for \ion{H}{ii}  regions (filled circles), young PNe (filled squares), old PNe (filled triangles). O, N, Ne, S and Ar observational abundances of \ion{H}{ii}  regions and PNe are from Table 4 (this work). 
 C/O and Cl/O ratios were adopted from Peimbert et al. (2005). Data shown in the Fe/O graph (lower right) correspond to
the average Fe/H ratio of two A-type supergiants reported by Venn et al. (2001) and the O/H value
of this work.
}
\end{center}
\end{figure*}

\subsubsection{ The results}

Our model, presented in Fig. 3,  was tailored to reproduce the current O/H value as given by \ion{H}{ii}  regions  (values from Table 4).  
It  is equivalent to Model 7L by CCP, except that they assumed a KTG IMF with a mass upper-limit equal to 60 M$_\odot$,
while  our model needs  to reduced this limit to 40 M$_\odot$, in order to match a lower current O/H.

In Fig. 3,  the 12\,+\,log\,(O/H) and 12\,+\,log\,(Ar/H) evolution with time is shown,
 both elements behave very similarly with time. 
A fast enrichment is predicted during the first 2 Gyr,
 because Ar and O are produced mainly by MS. The enrichment is
followed by a light fall (sort of plateau) up to 6 Gyr, and then a second fast enrichment occurs, 
particularly accelerated in the last couple of Gyr, imitating the time dependent of SFH.
The observational data for different objects 
(\ion{H}{ii} regions, young and old PNe, and A-type stars) are also shown in Fig. 3.

We find that the predicted evolution of O/H and Ar/H for $t\,<\,7$ Gyr
agrees with the average abundances of old PNe, 
considered to be 6$\pm$3 Gyr old. This epoch coincides with the beginning of the second fast enrichment.
Moreover, the recent evolution of O/H and Ar/H matches
 with the average abundances of young PNe,
assumed to be 2$\pm$1 Gyr old (filled square in Fig. 3).

The observational point (filled square) seems slightly deviated from the model line, 
but taking into account the large uncertainty in PN age, we consider that the model reproduces this point. 
Since O/H and Ar/H evolution depends mainly on star formation rate and the agreement with observations is pretty good,
we conclude the SFH inferred by photometric properties and assumed in this work, 
truly represent the star formation history in NGC\,6822.

Since our model assumes that most of LIMS formed during the whole evolution of the galaxy do not produce O
and that model reproduces well  both, O and Ar, abundances in PNe, then
 it is apparent that O  has not been perturbed in PN progenitors of NGC\,6822. 
Therefore, there is no evidence for significant third dredge-up affecting O abundance at this metallicity.

In Fig. 3 we also present  log\,(X$_{\rm i}$/O) vs. 12\,+\,log\,(O/H) behavior 
for different elements X$_{\rm i}$ (C, N, Ne, S, Cl, and Fe) obtained from our model. 
We have chosen to represent elemental abundance ratio as a function of O  
because this is a common way to display abundances in photoionized nebulae and in evolution modelling. 
For this figure, we have taken log\,(C/O)\,=\,$-0.31\pm$0.13 and log\,(Cl/O)\,=\,$-3.71\pm$0.10  from Peimbert et al. (2005), while log\,(Fe/H) \,=\,$-5.04\pm0.20$  is from Venn et al. (2001). This value combined with our O/H value from \ion{H}{ii} regions, gives a present-day  abundance ratio of log (Fe/O) = $-1.1\pm$0.5.
The figures show that for all the elements, but N and C, the X$_{\rm i}$/O vs. O/H evolutions predicted by the model match very well 
with \ion{H}{ii} region and PN data within one $\sigma$ errors.

The predicted present-day C/O and N/O ratios are $\sim$\,0.4 and 1.0 dex, respectively, higher than the observed values. 
The C/O discrepancy is  complex because the value we are using was computed by Peimbert et al. (2005) from recombination lines (RL), while in this work we are using abundances from collisionally excited lines (CEL). 
However Peimbert et al. (2005) argue that C/O ratio does not depends on the temperature structure of the nebula, and thus   
(C/O)$_{\rm CEL}$ = (C/O)$_{\rm RL}$, in which case the discrepancy should be due to other causes.
 Moreover, the dust correction considered by Peimbert et al. does not change their C/O value,
because that correction increases the C/H and O/H ratios by 0.10 and 0.08 dex, respectively.

Part of the problem with both elements is due to the fact that we have reduced the mass-upper limit of
 KTG IMF to 40 M$_\odot$, in order to match the low current O/H obtained for \ion{H}{ii} regions. As C and N are mainly produced by LIMS,
  these elements are not reduced equivalently. In addition, at least for the case of N,  the yields by Karakas \& Lattanzio (2007) 
 are higher than those provided by other authors, mainly for stars with M$\>$ 5\,M$_\odot$ and intermediate metallicity ($Z$=0.008), thus providing a higher N abundance. This problem will be discussed  deeply in Hern\'andez-Mart{\'\i}nez et al. (2009, in preparation) where a whole set of models will be presented.

\section{ Conclusions}

From MOS spectroscopy performed with 8-m class telescopes we derived chemical compositions for 11 PNe and 1 \ion{H}{ii} region in NGC\,6822. From this, our results are the following

a) Combining our  \ion{H}{ii} region abundance with other well determined \ion{H}{ii} region values from the literature, we confirm that the present  ISM is chemically homogeneous, at least in the central 2 kpc of NGC\,6822. The present average oxygen abundance obtained is $<$12 + log (O/H)$>$= 8.06$\pm$0.04

b) PNe abundances spray in a larger range of abundances than \ion{H}{ii} regions. According to their Ar/H abundances, we identify a young PN population 
with abundances similar to \ion{H}{ii} regions and an older population with abundances a factor of two lower.

c) All  PNe present  N-enrichment relative to \ion{H}{ii} regions. Two of them are extremely N-rich (Type I PNe), and a third one, from the old population also present N-enrichment compatible with a Type I PN.

d) Ne and O abundances in our objects, follow a very tight linear correlation with slope near 1. Some young non Type I PNe seem to have suffered slight O and Ne-enrichment, previously to the nebular ejection, showing abundance values slightly higher than \ion{H}{ii} regions. This effect however could be attributed to other causes, for example,  dust in \ion{H}{ii} regions. Interestingly, the old PN population, with 12\,+\,log (O/H) $\leq$ 7.8, do not seem to be O-enriched as compared with the predictions of our chemical evolution model. Thus, the third dredge-up of O  did not occurred in the PNe of NGC\,6822. This would indicate  that for a significant O-enrichement, 
 a  lower metallicity is necessary.

e) The pattern of PNe  chemical abundances in NGC\,6822 follows closely that of the SMC, where also a mixture of populations is found.

For the first time, thanks to PN abundance determinations,
a complete evolution model of NGC 6822 is obtained.
The model, tailored to match the present O/H abundance ratio determined in this work,
matches the abundance ratios of heavy elements, produced mainly by massive stars,
and predicts well the behavior of most of elements observed in the young and old PN populations. The model however needs some refinements
in order to match C and N abundances, mainly produce by LIMS.
 
Based on a galactic formation given by a $\Lambda$CDM cosmology
and  a star formation history increasing and bursting during the whole evolution of the galaxy,
we infer the following characteristics for the chemical history of NGC 6822:

i) an important gas-mass lost  occurred during the first 5.3 Gyr,

ii) no star higher than 40 M$_\odot$ was formed, and

iii) One percent of all 3-15 M$_\odot$ stars became binary systems progenitors to SNIa.

\begin{acknowledgements}
Invaluable comments and support by  Manuel Peimbert and Michael Richer are deeply appreciated.  M. Pe\~na is grateful to DAS, Universidad de Chile, for hospitality during a sabbatical stay when part of this work was performed.  L. H.-M. benefited from the hospitality of the Departamento de Astronom{\'\i}a, Universidad de Chile for this work.  L. H.-M. received a scholarship from  CONACYT-M\'exico and DGAPA-UNAM. J. G.-R. is supported by a postdoctoral grant from UNAM. 
 M. P. gratefully acknowledges financial support from FONDAP-Chile and DGAPA-UNAM. This work received financial support from CONACYT-M\'exico (grants \#43121, \#46904 and \#60354) and   DGAPA-UNAM (grants IN-114805 and IN-112708).
  \end{acknowledgements}

\clearpage
\setcounter{table}{1}
\begin{landscape}
\begin{table}
%\begin{center}
{\footnotesize
\caption{Dereddened line intensities, relative to H$\beta$=1.0}
\begin{tabular}{llrrrrrrrrrrrrrrr}
%\longtabL{2}{
%\begin{landscape}
%\begin{longtable}{llrrrrrrrrrrrrrrr}
%\begin{center}
%{\footnotesize
\hline \hline
$\lambda$&ion&f($\lambda$)&PN\,4 &\%&PN\,5 &\%&PN\,6 &\%&PN\,7 &\%&PN\,10 &\%&PN\,12&\%&PN\,14 &\%\\
\hline 
%\endfirsthead
%\caption{continued.}\\
%\hline \hline
%$\lambda$&ion&f($\lambda$)&PN\,4 &\%&PN\,5 &\%&PN\,6 &\%&PN\,7 &\%&PN\,10 &\%&PN\,12&\%&PN\,14 &\% \\
%\hline
% \endhead
3426&[\ion{Ne}{v}]&0.342&0.227&30\\ 
3727&[\ion{O}{ii}]&0.255&0.521&10&&&0.445&15&&&$<$0.043&&0.197&20\\
3749&\ion{O}{ii}&0.253&0.048&20&&&0.076&40\\
%3760&\ion{O}{iii}&0.252&0.014&20&&&&\\
3771&H11&0.250&0.036&20&&&0.053&40&&&&&0.067&30&&\\
3798&H10&0.245&0.041&15&&&0.071&40\\
3835&H9&0.236&0.065&15&&&0.075&20&&&0.033&50&0.066&30&\\
3869&[\ion{Ne}{iii}]&0.220&0.511&5&&&0.460&8&&&0.281&10&0.543&8\\
3889&H8+\ion{He}{i}&0.223&0.222&5&&&0.178&15&&&0.101&15&0.180&10\\
3969&H7+[\ion{Ne}{iii}]&0.203&0.294&5&&&0.314&10&&&0.208&10&0.326&8\\
4025&\ion{He}{i}+\ion{He}{ii}&0.190&0.020&15&&&0.044&20&&&0.057&40\\
4069&[\ion{S}{ii}]&0.187&0.031&15&&&0.027&30&&&0.036&50&0.027&30&0.038&15\\
4101&H$\delta$&0.182&0.263&8&&&0.265&8& & &0.293&10&0.263&8&0.277&5\\
4227&[\ion{Fe}{v}]&0.150&0.024&20\\
%4238&\ion{N}{ii}?&0.140&0.005&50\\
4340&H$\gamma$&0.125&0.474&8&0.470&20&0.469&8&0.470&10&0.478&8&0.466&8&0.464&5\\
4363&[\ion{O}{iii}]&0.121&0.188&12&0.236&40&0.097&12&$\leq$\,0.106&&0.125&10&0.138&10&0.305&8\\
4471& \ion{He}{i}&0.078&0.040&10&&&0.059&15&$\leq$\,0.058&&0.069&20&0.036&20&0.035&20\\
4541&\ion{He}{ii}&0.076&0.013&30&&&&&&&&&&&0.025&20\\
4640&\ion{N}{iii}&0.073&0.007&50&&&0.009&50&&&&&&&0.052&18\\
%4658&[\ion{Fe}{iii}]&0.072&0.011&30&&\\
4686&\ion{He}{ii}&0.070&0.221&10&0.523&10&0.010&50&$<$\,0.100&&$<$\,0.026&&0.235&13&0.586&5\\
4711&[\ion{Ar}{iv}]&0.039&0.030&15&&&0.015&40&&&&&0.034&30&0.028&25\\
4740&[\ion{Ar}{iv}]&0.030&0.027&15&&&0.008&40&&&&&0.030&30&0.044&20\\
4861&H$\beta$&0.000&1.000&5&1.000&3&1.000&5&1.000& 8&1.000&3&1.000&5&1.000&1\\
4922&\ion{He}{i}&-0.014&0.009&20&0.040&20&0.012&30\\
4959&[\ion{O}{iii}]&-0.023&2.276&3&1.633&3&2.492&3&2.330&5&1.419&3&3.346&5&3.602&1\\
5007&[\ion{O}{iii}]&-0.033&6.695&3&5.169&3&7.589&3&7.170&5&4.361&2&10.047&5&10.666&1\\
5016&\ion{He}{i}&-0.036&0.022&15&&&0.028&15&&&0.019&50&0.022&30&\\
5146&[\ion{Fe}{vi}]&-0.050&0.014&30\\
5200&[N I]&-0.083&&&&&0.004&50&&&&&&&0.097&10\\
5411&\ion{He}{ii}&-0.115&0.019&15&&&&&&&&&0.046&50&0.054&12\\
%5518&[\ion{Cl}{iii}]&-0.143&$<$ 0.007 &&&&&&&&&&&&$<$0.0080\\
%5538&[\ion{Cl}{iii}]&-0.152&$<$  0.007 &&&&&&&&&&&&$<$0.0080\\
5755&[\ion{N}{ii}]&-0.191&0.012&30&0.115&15&0.004&50&&&&&&&0.239&5\\
5876&\ion{He}{i}&-0.208&0.123&8&0.085&20&0.146&8&0.170&20&0.168&10&0.145&10&0.110&10\\
%6086&??&-0.240&0.0182&20\\
\hline \hline\\
\label{tab:tttt}
%\end{center}
%\end{longtable}
%\end{landscape}
\end{tabular}
}
%\end{center}
\end{table}
\end{landscape}

\clearpage
\setcounter{table}{1}
\begin{landscape}
\begin{table}
%\begin{center}
{\footnotesize
\caption{(continued )Dereddened line intensities, relative to H$\beta$=1.0}
\begin{tabular}{llrrrrrrrrrrrrrrr}
%\longtabL{2}{
%\begin{landscape}
%\begin{longtable}{llrrrrrrrrrrrrrrr}
%\begin{center}
%{\footnotesize
\hline \hline
$\lambda$&ion&f($\lambda$)&PN\,4 &\%&PN\,5 &\%&PN\,6 &\%&PN\,7 &\%&PN\,10 &\%&PN\,12&\%&PN\,14 &\%\\
\hline 

6300&[\ion{O}{i}]&-0.284&0.031&15&0.254&15&&&0.058&40&0.014&30&0.016&20&0.278&5\\
6312&[\ion{S}{iii}]&-0.286&0.025&10&0.034&40&0.009&40&0.010&40&$< $\,0.013&&0.017&20&0.012&50\\
6364&[\ion{O}{i}]&-0.294&0.010&15&0.028&40&&&0.023&40&&&&&0.091&13\\
6548&[\ion{N}{ii}]&-0.300&0.124&8&1.080&5&0.059&12&0.097&20&0.018&40&0.019&25&1.382&1\\
6563&H$\alpha$&-0.322&2.860&2&2.750&3&2.858&2&2.860&5&2.749&2&2.862&3&2.749&1\\
6583&[\ion{N}{ii}]&-0.325&0.351&8&3.203&3&0.158&8&0.284&15&0.045&25&0.059&20&4.180&1\\
6678&\ion{He}{i}&-0.339&0.029&10&$<$\,0.028&&0.040&12&0.057&30&0.035&25&0.035&25&0.027&25\\
6716&[\ion{S}{ii}]&-0.343&0.036&10&0.26&15&0.017&15&0.014&40&$<$\,0.013&&0.004&50&0.021&28\\
6730&[\ion{S}{ii}]&-0.344&0.054&10&0.27&15&0.029&12&0.021&40&$<$\,0.013&&0.009&30&0.036&20\\
7006&[\ion{Ar}{v}]&-0.375&0.006&30&$<$\,0.017&&&&&&&&0.008&30&0.026&25\\
7065&\ion{He}{i}&-0.383&0.063&8&$<$\,0.028&&0.094&8&0.131&20&0.118&15&0.054&20&0.060&15\\
7136&[\ion{Ar}{iii}]&-0.391&0.065&8&0.114&20&0.056&8&0.099&30&0.026&30&0.053&20&0.080&15\\
7281&\ion{He}{i}&-0.406&0.013&15&&&0.015&15&\\
7319&[\ion{O}{ii}]&-0.410&0.045&10&0.035&30&0.036&10&0.183&15&0.034&25&$<$\,0.018&25&0.139&10\\
7331&[\ion{O}{ii}]&-0.412&0.035&10&&&0.030&10&0.151&20&0.033&25&&&0.116&10\\
7751&[\ion{Ar}{iii}]&-0.421&0.017&15&&&0.017&15&0.036&40&&&0.029&25&0.025&25\\
\hline
log F(Hb)&&&-14.804&&-15.300:& &-15.090&&-14.950:& &-15.184&&-15.312&&-15.011&\\
c(Hb)&&&0.52&&---&&0.58&&---&&0.80&&0.00&&0.40&\\
\hline \hline\\
\label{tab:tttt2}
%\end{center}
%\end{longtable}
%\end{landscape}
\end{tabular}
}
%\end{center}
\end{table}
\end{landscape}

\clearpage
\setcounter{table}{1}
\begin{landscape}
\begin{table}
%\begin{center}
{\footnotesize
\caption{(continued) Dereddened line intensities, relative to H$\beta$=1.0}
\begin{tabular}{llrrrrrrrrrrrrrrr}
%{\footnotesize
%\rotate{
\hline \hline
$\lambda$&ion&f($\lambda$)&PN\,16 &\%&PN\,18 &\%&PN\,19 & \%&PN\,21 &\% &\ion{H}{ii}\,15&\% &\ion{H}{ii}\,9&\% &\ion{H}{ii}\,04&\%\\
\hline 
%\endfirsthead
%\caption{continued.}\\
%\hline \hline
%$\lambda$&ion&f($\lambda$)&PN\,16 &\%&PN\,18 &\%&PN\,19 & \%&PN\,21 &\% &\ion{H}{ii}\,15&\% &\ion{H}{ii}\,9&\% &\ion{H}{ii}\,04&\%\\
%\hline
 %\endhead
%3426&[\ion{Ne}{v}]&0.342&&&&&&&&&&&&&Gemini&\\
3727&[\ion{O}{ii}]&0.255&0.156&20&&&&&0.468&30&2.128&2&4.309&10&0.445&30\\
3749&\ion{O}{ii} &0.253&&&&&&&&&0.039 & 20\\
%3760&\ion{O}{iii}I&0.252&&&&&&\\
3771&H11&0.250&0.053&30&&&&&&&0.046&5\\
3798&H10&0.245&0.068&30&&&&&&&0.060&5\\
3835&H9&0.236&0.034&35&&&&&&&0.080&5\\
3869&[\ion{Ne}{iii}]&0.220&0.831&10&0.308&20&&&0.425&20&0.254&3&&&0.044&60\\
3889&H8+\ion{He}{i}&0.223&0.162&20&&&&&0.229&30&0.196&3\\
3969&H7+[\ion{Ne}{iii}]&0.203&0.387&15&0.242&20&0.196&30&0.200&20&0.234&3&&&0.066&60\\
4025&\ion{He}{i}+\ion{He}{ii}&0.190&0.026&30&&&&&&&0.021&10\\
4069&[\ion{S}{ii}]&0.187&0.056&25&&&&&&&0.013&12&0.085&30\\
4101&H$\delta$&0.182&0.248&10&0.164&30&0.247&25&0.260&20&0.258&3&0.233&20&0.245&12\\
%4143&\ion{He}{i}&0.175&&&&&&&&&&&\\
%4200&\ion{He}{ii}&0.160&&&&&&&&&&\\
%4227&[Fe V]&0.150&&&&&&&&&\\
%4238&N II?&0.140&&&&&&&&&&\\
4340&H$\gamma$&0.125&0.470&8&0.470&15&0.441&10&0.460&10&0.472&3&0.466&10&0.474&8\\
4363&[\ion{O}{iii} I]&0.121&0.206&10&0.084&15&0.129&25&0.140&15&0.042&5&&&$<$\,0.01\\
4471& \ion{He}{i}&0.078&0.057&20&&&&&&&0.041& 8\\
%4541&\ion{He}{ii}&0.076&\\
%4640&N III&0.073&\\
4658&[\ion{Fe}{iii}]&0.072&&&&&&&&&0.007&20\\
4686&\ion{He}{ii}&0.070&0.303&10&0.741&5&0.166&15&0.027&&---&\\
4711&[\ion{Ar}{iv}]&0.039&0.044&20&&&$<$\,0.045&&&&0.006&20\\
4740&[\ion{Ar}{iv}]&0.030&0.049&20&&&$<$\,0.045&&&&---\\
4861&H$\beta$&0.000&1.000&3&1.000&5&1.000&2&1.000&5&1.000&1&1.000&5&1.000&5\\
4922&\ion{He}{i}&-0.014&&&&&&&&&0.009&10\\
4959&[\ion{O}{iii}]&-0.023&3.835&3&0.926&5&1.139&2&2.120&5&1.360&1&&&0.075&15\\
5007&[\ion{O}{iii}]&-0.033&11.806&2&2.725&5&3.266&1&6.091&3&3.604&1&0.051&20&0.221&10\\
5016&\ion{He}{i}&-0.036&&&&&&&&&0.020& 10\\
%5146&[Fe VI]&-0.050&&&&&&&&&\\
5200&[\ion{N}{i}]&-0.083&&&&&&&&&0.005&20&&&0.019&30\\
5411&\ion{He}{ii}&-0.115&0.029&25&0.098&30&&&&&\\
5518&[\ion{Cl}{iii}]&-0.143&&&&&&&&&0.004&20\\
5538&[\ion{Cl}{iii}]&-0.152&&&&&&&&&0.003&30\\
5755&[\ion{N}{ii}]&-0.191&&&&&&&&&0.002&30\\
5876&\ion{He}{i}&-0.208&0.197&8&$<$\,0.096&&0.146&35&0.128&20&0.106&10&0.060&40&0.020&20\\
%6086&??&-0.240&&&&&&&&\\
\hline \hline\\
\label{tab:tttt3}
%\end{center}
%\end{longtable}
%\end{landscape}
\end{tabular}
}
%\end{center}
\end{table}
\end{landscape}

\clearpage
\setcounter{table}{1}
\begin{landscape}
\begin{table}
%\begin{center}
{\footnotesize
\caption{(continued) Dereddened line intensities, relative to H$\beta$=1.0}
\begin{tabular}{llrrrrrrrrrrrrrrr}
%{\footnotesize
%\rotate{
\hline \hline
$\lambda$&ion&f($\lambda$)&PN\,16 &\%&PN\,18 &\%&PN\,19 & \%&PN\,21 &\% &\ion{H}{ii}\,15&\% &\ion{H}{ii}\,9&\% &\ion{H}{ii}\,04&\%\\
\hline 
6300&[\ion{O}{i}]&-0.284&0.052&15&&&0.111&50&&&0.014&10&&&0.010&40\\
6312&[\ion{S}{iii}]&-0.286&0.012&25&&&0.035&50&&&0.019&10&&&0.008&40\\
6364&[\ion{O}{i}]&-0.294&0.023&20&&&&&&&0.006&12\\
6548&[\ion{N}{ii}]&-0.300&0.065&12&0.186&15&0.136&20&0.032&30&0.030&8&0.103&40\\
6563&H$\alpha$&-0.322&2.857&2&2.860&5&2.80&2&2.725&5&---&sat.&2.888&5\\
6583&[\ion{N}{ii}]&-0.325&0.164&0&0.367&15&0.252&15&0.114&15&0.091&5&0.247&10\\
6678&\ion{He}{i}&-0.339&0.032&15&&&&&0.059&20&0.028&8\\
6716&[\ion{S}{ii}]&-0.343&0.016&20&0.063&30&0.603&10&0.067&20&0.092&8&0.456&8\\
6730&[\ion{S}{ii}]&-0.344&0.018&20&0.074&30&0.401&10&0.054&20&0.074&8&0.318&8\\
%7006&[Ar V]&-0.375&&&&&&&&&&\\
7065&\ion{He}{i}&-0.383&0.093&10&&&&&0.063&20&0.023&10\\
7136&[\ion{Ar}{iii}]&-0.391&0.059&15&&&0.088&30&0.071&20&0.096&8\\
7281&\ion{He}{i}&-0.406&&&&&&&&&0.005&12\\
7319&[\ion{O}{ii}]&-0.410&0.053&15&&&0.074&30&0.036&30&0.021&10\\
7331&[\ion{O}{ii}]&-0.412&0.042&15&&&0.139&30&0.038&30&0.023&10\\
%7751&[Ar III]&-0.421&&&&&&&&&&\\
\hline
log F(Hb)&&&-15.406&&-16.059&&-16.073&&-15.936&&-14.165&&-15.33:&&-14.342\\
c(Hb)&&&0.35&&0.00&&0.35&&0.00&&0.84&&0.44:&&0.40\\
\hline \hline\\
\label{tab:tttt4}
%\end{center}
%\end{longtable}
%\end{landscape}
\end{tabular}
}
%\end{center}
\end{table}
\end{landscape}

%\section{Appendix}

\begin{thebibliography}{}
\bibitem[]{allen}Allen, C., Carigi, L., \& Peimbert, M. 1998, \apj, 494, 247
\bibitem[]{Bauman} Bauman, R. P., Porter, R. L., Ferland, G. J., \& MacAdam, K. B. 2005, \apj, 628, 541
\bibitem[]{} Brandenburg, H. J., \& Skillman, E. D. 1998, BAAS, 30, 1354
\bibitem[]{carigi99} Carigi, L, Colin, P., \&  Peimbert, M. 1999, \apj, 514, 787
\bibitem[]{carigi} Carigi, L., Colin, P., \& Peimbert, M. 2006, ApJ, 644, 924 (CCP)
\bibitem[]{carher} Carigi, L., \& Hernandez, X. 2008, MNRAS, 390, 582
\bibitem[]{costa2000} Costa, R. D. D., de Freitas Pacheco, J. A., \& Idiart, T. P. 2000, A\&AS, 145, 467
\bibitem[deBlok2000]{blok00} de Blok, W. G. J., \&  Walter, F. 2000,  \apj, 537, 95
\bibitem[deBlok2006]{blok06} de Blok, W. G. J., \&  Walter, F. 2006, \aj, 131, 343
\bibitem[]{demers06} Demers, S., Batinelli, P., \&  Artigau, E. 2006,  A\&A,  456, 905
\bibitem[]{dennefeld} Dennefeld, M. 1989, Recent development of Magellanic Clouds research, eds. K. S: de Boer, F. Spite, \& G. Stasinska, 107
%\bibitem[]{demers06b} Demers, S., Batinelli, P.,  Kunkel, W.E., 2006b, \apj, 636, L85
\bibitem[]{dufour} Dufour, R. J., \&   Talent, D. L.  1980, \apj, 235, 22
%\bibitem[]{} Fitzpatrick, E. L. 1999, PASP, 111, 63
\bibitem[]{} Fisher, J. R., \& Tully, R. B. 1979, AJ, 84, 62 
%is not quoted\bibitem[]{} Chieffi, A.,  Limongi, M. 2002, ApJ, 577, 281
%\bibitem[]{ford} Ford, H., Peng, E., \& Freeman, K. 2002, The Dynamics, Structure \& History of Galaxies, ASP Conference Series, 273, 41
\bibitem[]{} Franco, I., \&  Carigi, L. 2008, RMAA, 44, 311
%\bibitem[]{}Gallart, C., Aparicio, A., Bertelli, G., \& Chiosi, C.  1996, AJ, 112, 1950
\bibitem[]{garnet99} Garnett, D. 1999, New Views of the Magellanic Clouds, IAU Symp. 190, eds. Y.-H. Chu, N. B. Suntzeff, J. E. Hesser \& D. A. Bohlender, 266
\bibitem[gieren]{gieren06} Gieren, W.,  Pietrzy\'nski, G., Nalewajko, K., et al. 2006, \apj, 647, 1056
\bibitem[]{} Grevesse, N., Asplund, M., \& Sauval, A. J. 2007, SSRv, 130, 105
%is not quoted\bibitem[]{} Greggio, L., Renzini, A. 1983, A\&A, 118, 217
%\bibitem[]{} Hanes, D. A, \& Whittaker, D. G. 1987, \aj, 94, 906
\bibitem[]{henry89} Henry, R. B. C  1989, MNRAS, 241, 453
\bibitem[]{henry90} Henry, R. B. C. 1990, \apj, 356, 229
\bibitem[]{hernandez} Hern\'andez-Mart{\'\i}nez, L., \& Pe\~na, M. 2009, A\&A,  495, 447 (Paper I)
%\bibitem[]{herrmann08} Herrmann, K. A., Ciardullo R., Feldmeir, J. J., \& Vinciguerra, M.  2008, astro-ph/0805-1074
\bibitem[]{} Herwig, F. 2004, ApJS, 155,651
\bibitem[]{hidalgo01} Hidalgo-G\'amez, A. M., Masegosa, J., \& Olofsson, K. 2001, A\&A, 369, 797
\bibitem[]{} Hirschi, R. 2007, \aap, 461, 571
\bibitem[]{} Hirschi, R., Meynet, G., \&  Maeder, A. 2005, \aap, 433, 1013
%\bibitem[Hodge et al. 1988]{hodge88} Hodge, P., Kennicutt, R. C. Jr., \& Lee, M. G. 1988, \pasp, 100, 917
\bibitem[]{hodge 1991} Hodge, P., Smith, T., Eskridge, P., MacGilivray, H., \& Beard, S. 1991, \apj, 379, 621
%\bibitem[Jacoby 1989]{jacob89}Jacoby, G. H. 1989, \apj, 339, 39
%\bibitem[]{jacoby-demarco} Jacoby, G. H, \& De Marco, O. 2002, AJ, 123, 269
\bibitem[]{} Karakas, A., \&  Lattanzio, J. C. 2007, PASA  24, 103 
\bibitem[]{killen} Killen, R. M., \&  Dufour, R. J. 1982, PASP, 94, 444
\bibitem[]{} Kingdon, J., \& Ferland, G. J. 1995, \apj, 442, 714
\bibitem[]{kingsburgh} Kingsburgh, R., \&  Barlow, M. 1994, MNRAS, 271, 257 
\bibitem[]{kniazev05} Kniazev, A. Y., Grebel, E. K., Pustilnik, S. A., Pramskij, A. G., \& Zucker, D. B. 2005, \aj, 130, 1558
\bibitem[]{} Kroupa, P., Tout, C. A., \& Gilmore, G. 1993, MNRAS, 262, 545 (KTG)
\bibitem[]{lee} Lee, H.,  Skillman, E. D., \& Venn, K. A. 2006, \apj, 642, 813
\bibitem[Leisy et al. 2005]{Leisy05}Leisy, P., Corradi, R. L. M., Magrini, L., et al.  2005,  A\&A, 436, 437
\bibitem[]{LD96} Leisy, P., \& Dennefeld, M. 1996, A\&ASS, 116, 95
\bibitem[]{LD06} Leisy, P., \& Dennefeld, M. 2006, A\&A, 456, 451 

%\bibitem[]{2000} Magrini, L., Corradi, R. L. M., Mampaso, A., \& Perinotto, M. 2000, A\&A, 355, 713
%\bibitem[2005]{martins05} Martins, F., Schaerer, D., \& Hillier, D. J.  2005, A\&A, 436, 1049
%\bibitem[]{} Maeder, A. 1992, A\&A, 264, 105
\bibitem[]{magrini09} Magrini, L., Stanghellini, L., \& Villaver, E. 2009, \apj, 696, 729 
\bibitem[]{magrini05} Magrini, L., Leisy, P., Corradi, R. L. M., Perinotto, M., Mampaso, A., \&  V{\'\i}lchez, J. M. 2005, A\&A, 443, 115
\bibitem[]{} Mannucci 2007, Multifrequency behaviour of high energy cosmic sources,
 eds. F. Giovannelli \& L. Sabau-Graziati, ChJAA (arXiv:0708.0472)
\bibitem[]{} Marigo, P.  2001, \aap, 370, 194
\bibitem[]{} Massey, P., Armandroff, T. E., Pyke, R., Patel, K., \& Wilson, C. D. 1995, \aj, 110,  2715
\bibitem[]{mateo98} Mateo, M. 1998, \araa, 36, 435
\bibitem[]{} Meynet, G., \&  Maeder, A. 2002, A\&A, 390, 561
\bibitem[]{} Nomoto, K., Iwamoto, K., Nakasato, N.  et al. 1997, NuPhA, 621, 467
\bibitem[]{peimbert78} Peimbert, M. 1978, Planetary Nebulae, IAU Symp. 76, ed. Y. Terzian, p. 215  
\bibitem[]{peimbert85} Peimbert, M. 1985, RevMexAA, 10, 125
\bibitem[]{peimbert05} Peimbert, A., Peimbert, M., \&  Ruiz, M. T. 2005, \apj, 634, 1056
\bibitem[]{pena07} Pe\~na, M., Richer, M. G., \&  Stasinska, G. 2007, A\&A, 466, 75
%is not quoted\bibitem[]{} Portinari, L., Chiosi, C.,  Bressan, A. 1998, A\&A, 334, 505
\bibitem[]{}  Porter, R. L., Ferland, G. J., \& MacAdam, K. B. 2007, ApJ, 657, 327 
\bibitem[]{richer05} Richer, M. G., \&  McCall, M. 1995, \apj, 445, 642
\bibitem[]{richer07} Richer, M. G., \&  McCall, M. 2007, \apj, 658, 328
\bibitem[]{seaton79} Seaton, M. 1979, MNRAS, 185, 5
\bibitem[]{} Shaw, R. A., \&  Dufour, R. J.  1995, PASP, 107, 896
\bibitem[]{} Stasi\'nska, G., Richer, M., \& McCall, M. 1998, A\&A, 336, 667
\bibitem[]{} Storey, P. J., \&  Hummer, D G. 1995, MNRAS, 272, 41
\bibitem[]{venn} Venn, K. A.,  Lennon, D. J., Kaufer, A., et al. 2001, \apj, 625, 754
\bibitem[]{wang} Wang, W., \&  Liu, X.-W., 2008, MNRAS, 389, L33
%\bibitem[]{}  Weldrake, D. T. F., de Blok, W. J. G.,  \& Walter, F., 2003, MNRAS, 340, 12
\bibitem[]{} Woosley, S. E., \&   Weaver, T. A. 1995, ApJS, 101, 181
\end{thebibliography}
\end{document}